\documentclass[twocolumn]{aastex62}
\pdfoutput=1 
\usepackage{amsmath,amstext}
\usepackage[T1]{fontenc}
\usepackage[figure,figure*]{hypcap}
\graphicspath{{./}{figures/}}

\received{?}
\revised{?}
\accepted{?}
\submitjournal{AAS}

\shorttitle{Homogeneous Analysis of Hot Earths: Masses, Sizes, and Compositions}
\shortauthors{Dai et al.}

\begin{document}

\title{Homogeneous Analysis of Hot Earths: Masses, Sizes, and Compositions}

\author[0000-0002-8958-0683]{Fei Dai}
\affiliation{Department of Physics and Kavli Institute for Astrophysics and Space Research,\\Massachusetts Institute of Technology, Cambridge, MA, 02139, USA}
\affiliation{Department of Astrophysical Sciences, Princeton University, 4 Ivy Lane, Princeton, NJ, 08544, USA}
\email{fd284@mit.edu}

\author[0000-0003-1298-9699]{Kento Masuda}
\affiliation{Department of Astrophysical Sciences, Princeton University, 4 Ivy Lane, Princeton, NJ, 08544, USA}
\affiliation{NASA Sagan Fellow}

\author[0000-0002-4265-047X]{Joshua N.\ Winn}
\affiliation{Department of Astrophysical Sciences, Princeton University, 4 Ivy Lane, Princeton, NJ, 08544, USA}

\author[0000-0003-1957-6635]{Li Zeng}
\affiliation{Department of Earth \& Planetary Sciences, Harvard University, 20 Oxford Street, Cambridge, MA, 02138, USA}


\begin{abstract}

Terrestrial planets have been found orbiting Sun-like stars with extremely short periods --- 
some as short as 4 hours. These ``ultra-short-period planets'' or ''hot Earths''
are so strongly irradiated that any initial H/He atmosphere has probably been lost to
photoevaporation. As such, the sample of hot Earths may give us a glimpse
at the rocky cores that are often enshrouded by thick H/He envelopes on
wider-orbiting planets.
However, the mass and radius measurements of hot Earths have been
derived from a hodgepodge of different modeling approaches, and include
several cases of contradictory results. Here, we perform a homogeneous
analysis of the complete sample of 11 known hot Earths with an insolation
exceeding 650 times that of the Earth. We combine all available data for each planet,
incorporate parallax information from {\it Gaia} to improve the
stellar and planetary parameters, and use Gaussian Process regression to account for correlated noise in the radial-velocity data. 
The homogeneous analysis leads to a smaller dispersion in the apparent composition
of hot Earths, although there does still appear to be some
intrinsic dispersion.
Most of the planets are consistent with an Earth-like composition (35\% iron
and 65\% rock), but two planets (K2-141b and K2-229b) show evidence for a higher
iron fraction, and one planet (55\,Cnc\,e) has either a very low iron fraction
or an envelope of low-density volatiles.
All of the planets are less massive than 8\,$M_\oplus$, despite the selection
bias towards more massive planets, suggesting that 8\,$M_\oplus$ is the
critical mass for runaway accretion.
\end{abstract}




\section{Introduction} \label{sec:intro}

Planets with orbital periods shorter than 1 day are known as
``ultra-short-period'' (USP) planets \citep{Winn2018}.
They occur around $\approx$0.5\% of Sun-like stars, and
their radii rarely exceed 2\,$R_\oplus$ \citep{Sanchis}. Prime examples include CoRoT-7b \citep{Leger} and Kepler-78b \citep{Sanchis2013}.
The orbits of USP planets are so small that they are often
within the dust sublimation radius \citep[$a/R_\star \sim 8$ for Sun-like stars,][]{Isella}, where the formation of Earth-mass cores was probably impossible.
This makes the formation of USP planets an interesting problem, similar to the longstanding
problem of the formation of hot Jupiters.
In fact, an early idea was that USP planets are the
cores of hot Jupiters which lost their
gaseous envelopes due to tidal disruption or atmospheric erosion \citep[see, e.g.,][]{Valsecchi}.
This idea is now disfavored both because of the limited atmospheric erosion rate \citep[e.g.][]{Lammer2009} and the fact that the host stars of hot Jupiters and USP planets have
different metallicity distributions \citep{Winn2017}. Instead, the USP planets
seem to be more closely related to the more abundant population of
planets discovered by the {\it Kepler} mission,
with sizes ranging up to 4\,$R_\oplus$ and orbital periods shorter than a few months
\citep{Sanchis,Lee,Petigura}.

The radius distribution of these {\it Kepler} planets is bimodal, with a dip in occurrence
for planets with sizes between 1.5 and 2.0\,$R_\oplus$ \citep{Fulton}. This feature
had been predicted, based on theories of the photoevaporation
of H/He planetary atmospheres \citep{Owen}.
In the same theories, photoevaporation should have gone to
completion for USP planets, removing any H/He envelope and exposing the rocky core.
Supporting this picture is the fact
that almost all USP planets have sizes below 2\,$R_\oplus$, on the small end of the
bimodal radius distribution. In the alternative explanation of \citet{Zeng2019} that attributed the bimodal radius distribution to a rock-ice transition, the USP planets are also exposed rocky worlds.
Thus, by studying USP planets, we may be learning
about the rocky cores that lie inside the lower-density planets with sizes
between 2 and 4\,$R_\oplus$.
Moreover, the extremely short periods of USP planets facilitate
the measurement of their masses through the Doppler technique.
This is because the velocity amplitude varies as $P^{-1/3}$ and
because there is usually a clean separation between the
short orbital period and the longer timescale of stellar rotation (one of
the timescales over which stellar activity leads to spurious Doppler signals).

In this Letter, we present a homogeneous analysis of the masses, sizes, and compositions of
a sample of strongly
irradiated planets, hoping to illuminate the origin and evolution of these
planets through knowledge of their possible range of compositions.
Are they stripped of H/He atmospheres, as photoevaporation theory would require?
Did they migrate from beyond the snow line, in which case a large complement
of water or other volatile elements might still be present?
Were their rocky mantles stripped by giant impacts, leading to an enhancement
in the iron fraction similar to that of Mercury \citep[][]{Benz+2008}?

We felt a homogeneous analysis was warranted because 
the results in the literature were reported by different groups
using different procedures to determine the stellar parameters and to mitigate the
effects of time-correlated noise in the radial-velocity data.
For example, K2-106b was reported by \citet{Guenther} to have an iron mass
fraction of $80^{+20}_{-30}\%$, while \citet{Sinukoff2017} concluded that the
composition was compatible with that of the Earth (35\% iron).
For this work, we combined all the previously published datasets, used the parallaxes from {\it Gaia} Data Release 2 \citep{2018A&A...616A...1G} to refine the stellar parameters, and
employed a Gaussian Process framework \citep{Dai2017} to disentangle
planetary radial velocity signals from correlated noise.  Previous works \citep[e.g.][]{Southworth2008,Southworth2009} have demonstrated the value of a homogeneous analysis when comparing the properties of exoplanetary systems reported by different groups.

\section{Sample Selection}

The criterion $P_{\rm orb}$<1 day for USP planets is convenient, but a quantity
more directly relevant to our questions is the bolometric flux $F_{\rm p}$ delivered to the
planet from the star.  Our sample was defined by the requirement $F_{\rm p}/F_\oplus > 650$,
where the somewhat arbitrary number 650 was taken from the study of the
radius/flux distribution by \citet{Lundkvist}.  
We selected ``hot Earths'' by requiring $R_{\rm p}/R_\oplus < 2$. This radius
cut filtered out hot Jupiters as well as the extraordinary ``hot Neptune''
NGTS-4b, with $M_{\rm p}/M_\oplus = 20.6 \pm 3.0$, $R_{\rm p}/R_\oplus = 3.18 \pm 0.26$, and $F/F_\oplus \approx 950$ \citep{West}.
We also restricted attention
to FGK dwarf stars by restricting $\log g>4$.
Table 1 summarizes the characteristics of the 11 hot Earths in our sample.

\section{Stellar Parameters}

To determine the stellar parameters, we used the same procedures as \citet{Dai2018}, described below.
 We had previously
validated these procedures using a sample of stars
for which the radius and mass have also been determined precisely using asteroseismology.
The input information was the parallax from {\it Gaia} Data Release~2 \citep{2018A&A...616A...1G}; 
the observed spectroscopic parameters $T_{\rm eff}$, $\log g$, and [Fe/H];
the $K$-band apparent magnitude; and stellar evolutionary models
computed with {\it Modules for Experiments in Stellar Astrophysics}
\citep[MESA;][]{Paxton2011}.

The $K$-magnitudes were taken from the Two Micron All Sky Survey \citep[2MASS;][]{2006AJ....131.1163S}.
We corrected for extinction $A_K$ 
using the value of $E(B-V)$ and extinction vectors from the dust map Bayestar17 \citep{2018MNRAS.478..651G}.
We adopted a $30\%$ fractional uncertainty for $A_K$, following \citet{Fulton2018}.
We adopted the distance estimates from \citet{2018AJ....156...58B} instead of using the parallaxes directly, although the difference was minor for our sample stars, which are within a few hundred parsecs.
 We adopted spectroscopic parameters reported in the literature (Table 1). We inflated the uncertainties on uncertainties of $T_{\rm eff}$ to 110~K, $\log g$ to 0.1, and [Fe/H] to 0.1 to reflect the systematic offsets that may exist between the different methods used to infer the spectroscopic parameters.
We then sampled the MESA Isochrones \& Stellar Tracks
\citep{2016ApJS..222....8D} with the priors on the spectroscopic parameters, the $K$-band magnitude, the extinctions and the Gaia distance using the {\tt isochrones} package by
\citet{2015ascl.soft03010M}. This Bayesian analysis yielded posterior distribution of the stellar mass, radius and the stellar mean densities.
Table 1 reports the results.

\section{Transit Analysis}

For the {\it Kepler} systems in our sample, we used the Pre-search Data Conditioning light curves available on the Mikulski Archive for Space
Telescopes website\footnote{\url{https://archive.stsci.edu}}.
For the {\it K2} systems, we downloaded the target pixel files and produced the light curves using the procedure described by \citet{Dai2017}. We used short-cadence light curve whenever possible.
For 55\,Cnc, we used the HST/STIS light curve presented by \citet{Bourrier}.
For CoRoT-7, we downloaded the light curves from the CoRoT N2 public archive \footnote{\url{http://idoc-CoRoT.ias.u-psud.fr}}.

Using from the transit ephemerides reported in the literature (Table 1), we isolated the transit
data. Only the data obtained within one full transit duration of the transit midpoint were retained;
hence, the total timespan of each transit segment was two full transit durations.
We inspected the data from each transit segment visually, to remove obviously bad data points.
We fitted the light curves using the {\tt Batman} package \citep{Kreidberg2015}.
We assumed the limb darkening law to be quadratic and placed Gaussian priors
on the coefficients with widths of 0.3 and central values based upon theoretical
stellar atmosphere models computed with {\tt EXOFAST}\footnote{\url{astroutils.astronomy.ohio-state.edu/exofast/limbdark.shtml}.} \citep{Eastman2013}.
The other transit parameters included the orbital
period $P_{\rm orb}$, the midtransit time ($T_{\rm c}$), the planet-to-star
radius ratio ($R_{\rm p}/R_\star$), the scaled orbital distance
($a/R_\star$), and the cosine of the orbital inclination ($\cos I$) for each transiting planets in the system.
 For systems with only long-cadence (29.4 min) data,  we computed the theoretical transit model 30 times at 1-min interval. We then averaged the 30 flux values before comparing it with the observed flux level to mimic the effect of finite integration time.

We created a phase-folded transit light curve for subsequent analysis.
Two of the systems display detectable transit-timing variations (TTV): WASP-47 and Kepler-10 \citep{Becker,Ofir}. For those systems, we inserted an additional step. We began by fitting a constant-period transit model using all the available data. Then, we used the best-fitting model as a template to derive individual transit times. The phase-folded light
curve was created based on the model including TTVs.

We fitted simultaneously for the parameters of all the transiting
planets in each system. We imposed a Gaussian prior on the stellar mean density (Section 2),
a log-uniform prior on $R_{\rm p}/R_\star$, and a uniform prior on $\cos I >0$.
We assumed the orbit of the hot Earth to be circular, based on the theoretical
expectation that the tidal circularization timescale is on the order of 10\,Myr \citep{Winn2018}. The outer planets were allowed to have eccentric orbits.
For more efficient sampling, we did not use the eccentricity $e$ and the argument of pericenter $\omega$ as parameters; instead, we used $\sqrt{e}\cos\omega$ and $\sqrt{e}\sin\omega$ which
have a lower covariance.
We sampled the posterior distribution using the Markov Chain Monte Carlo technique
implemented in the {\sc emcee} code \citep{emcee}.
Table 2 reports the results for the key parameter $R_{\rm p}/R_\star$.

\section{Radial Velocity Analysis}

We compiled all available radial velocity data sets for each planet (see Table 2).
For WASP-47, we removed from consideration the 29 data points that were based
on spectra obtained during transit, because we did not want to model
the Rossiter-McLaughlin effect.
We modeled the radial velocity measurements in a Gaussian Process framework
similar to that described by \citet{Dai2017}.
In brief, we adopted a quasi-periodic kernel:
\begin{equation}
\label{covar}
C_{ij} = h^2 \exp{\left[-\frac{(\Delta t_{ij})^2}{2\tau^2}-\Gamma \sin^2{\frac{\pi\Delta t_{ij}}{T}}\right]}+\left[\sigma_i^2+\sigma_{\text{jit}}^2\right]\delta_{ij}
\end{equation}
where $C_{ij}$ is the covariance matrix, $\delta_{ij}$ is the Kronecker delta,
$\sigma_{\rm jit}$ is the ``stellar jitter'' (any noise component beyond the
measurement uncertainty that
is uncorrelated in time), $h$ is the covariance amplitude,
$t_i$ is the time of individual observation,
$\Delta t_{ij} = t_i - t_j$, $\tau$ is the correlation timescale,
$T$ is the period of the covariance,
and $\Gamma$ dictates the relative importance of the
Gaussian and periodic components of the kernel.
We set the priors for the ``hyperparameters'' of this kernel ($\tau$, $\Gamma$, and $T$)
by analyzing the out-of-transit flux variation in the light curves. The underlying idea
is that the both the flux variations and the spurious Doppler shifts originate from surface inhomogeneities of the host star, but the light curve has a higher precision and better
time sampling than the radial-velocity data.

After fitting the light curves with a Gaussian process,
we used the posteriors for the hyperparameters as priors in the radial-velocity analysis.
The adopted likelihood function has the following form:
\begin{equation}
\label{likelihood}
\log{\mathcal{L}} =  -\frac{N}{2}\log{2\pi}-\frac{1}{2}\log{|\bf{C}|}-\frac{1}{2}\bf{r}^{\text{T}}\bf{C} ^{-\text{1}} \bf{r}
\end{equation}
where $N$ is the number of data points, $\bf{C}$ is the covariance matrix, and $\bf{r}$ is the difference between the observed radial velocity and
the calculated Keplerian radial velocity.
Again, we assumed that the hot Earths have circular orbits, and allowed for
nonzero eccentricity of the outer orbits.
We imposed Gaussian priors on $P_{\text{orb}}$ and $T_{\text{c}}$ for the transiting planets based on the analysis of Section 4. We imposed log-uniform priors on the
RV semiamplitudes $K$, the covariance amplitude $h$, and the
jitter parameters. We imposed uniform priors on the RV offset of each spectrograph, $\sqrt{e}\cos\omega$ ($[-1,1]$) and $\sqrt{e}\sin\omega$ ($[-1,1]$). Again, we sampled the posterior distribution using {\sc emcee} \citep{emcee}.

Table 2 gives the resulting mass determinations
for the hot Earths. Figure 1 shows the radial-velocity results for the
representative system K2-141.

\section{Discussion}

Figure 2 compares the results of our uniform analysis to those reported in the literature.
The overall impression is that we found a smaller dispersion in the implied
compositions of the hot Earths. We were also able to achieve more precise
results in some cases. This is partly thanks to the parallax information provided by Gaia,
which led to substantial reduction in the planetary radius uncertainty for systems that
had been analyzed prior to Gaia Data Release 2. For example, Kepler-78b was reported to have $R_p = 1.20\pm0.09 R_\oplus$ by \citet{Howard2013}, while
we found $R_p = 1.228^{+0.018}_{-0.019} R_\oplus$. In other cases, improvement was
achieved by combining RV datasets from different groups, and by using Gaussian Process regression.

To quantify the constraints on composition, we used a simple two-layer
model, in which each planet is assumed to have an iron core and a rocky (MgSiO$_3$) mantle, using the procedure described by \citet{Zeng2016}.
Table 2 gives the resulting constraints on the iron mass
fraction.
One noteworthy result is that
K2-106b does not appear to be as iron-rich as had been previously
thought \citep{Guenther}. Instead of 80\%, our analysis
gave an iron fraction
of 40$\pm23$\%, which resembles the results for the other
hot Earths.

\subsection{Are USPs really rocky?}

All in all, our uniform analysis not only tightened the constraints on the mass and radius of individual planets, but also led to a smaller dispersion of iron fractions
centered on an Earth-like composition. The mean core mass fraction is $26\%$ and the
standard deviation is 23\%. In a few cases, the posterior for the iron mass
fraction extends to negative values; such cases should be interpreted
as an indication that the planet is probably composed of lower-density material
(volatiles) in addition to rock and iron.

We tested to see whether the core mass fraction of the planets is correlated with any of the stellar or planetary parameters, including bolometric insolation, orbital period, planetary mass, planetary radius and host star metallicity. None of these parameters showed a statistically significant correlation with the core mass fraction ($p<5$\%).
In particular, the lack of correlation with bolometric insolation can be interpreted
as additional evidence that hot Earths are free of H/He envelopes, as photoevaporation theory would predict.
If the planet's mass budget had even a few percent of H/He, there would likely
be a reduction in radius with increasing irradiation \citep{Lopez2017}
similar to the pattern that is seen for mini-Neptunes (2 to 4\,$R_\oplus$) just outside
of the ``hot Neptune desert'' \citep{Fulton2018}. Mini-Neptunes have systematically smaller radii when the bolometric flux is high.

Although an atmosphere of hydrogen and helium probably cannot survive on a hot Earth, what
about an atmosphere with a higher mean molecular weight, composed
of water, carbon dioxide, or other volatiles?
Simulations by \citet{Lopez2017} suggest that volatile atmospheres can survive on hot Earths  over geological timescales. Any hot Earths with substantial volatile atmospheres
might show up in between the ``rock'' and ``water'' lines in Figure 2.
However, we found no cases in which a substantial volatile atmosphere
is required to explain the data.
 To quantify this result, we assumed that these planets have Earth-like cores of Fe-MgSiO$_3$ ratio of 3:7 on top of which is a variable amount of water envelope.
Table 2 gives the $2-\sigma$ upper limits on the water mass fraction.
The individual upper limits generally favors a water content less than 10 or 20\% by
mass. In this sense, the hot Earths are consistent with a formation scenario in which
the planetesimals were volatile-depleted, i.e., within the snowline.

The two planets that appear to depart most from an Earth-like composition (although with
modest statistical significance) 
are WASP-47 e and 55 Cnc e.
\citet{Vanderburg2017} also highlighted these two systems as peculiar.
They are the only systems known that have both USP planets and
close-in giant planets (in 4 and 15-day orbits, respectively).
The host stars are also the most metal-rich among the sample, with [Fe/H] of 0.3 and 0.4.
Combined with their lower planetary densities, these may be clues that
the hot Earths formed in a different way around WASP-47 and 55\,Cnc.
They might be cases of disk migration from beyond the snowline.
This would explain the presence of gas giants, since it is easier
to achieved runaway gas accretion beyond the snowline, and it would
also explain why WASP-47\,e and 55\,Cnc\,e may have
volatile atmospheres.
It is worth noting WASP-47\,e and 55\,Cnc\,e are the largest hot Earths (by radius)
in our sample. Both have $R_{\rm p} > 1.6\,R_\oplus$, the approximate value
of the critical radius separating rocky and volatile-enhanced planets within
the larger sample of sub-Neptune planets \citep{Rogers}.
Furthermore, \citet{Angelo} have argued that 55\,Cnc\,e must possess some form of
thick atmosphere to explain the high heat redistribution efficiency that is indicated
in phase-curve observations \citep{Demory}.
Alternatively \citet{Dorn} proposed that these two planets are iron-poor and low in density because they are rich in calcium and aluminum whcih condensed earlier within
the protoplanetary disk.

\subsection{Threshold mass for runaway accretion}

As mentioned in the Introduction, there are several indications
that hot Earths are not a later evolutionary state of hot Jupiters,
but rather are drawn from the same basic population as the more abundant and longer-period sub-Neptune planets. This conclusion is also supported on more theoretical grounds. \citet{Ginzburg} studied the fate of Roche-Lobe-overflowing (RLO) hot Jupiters. They found that, depending on the planet's core mass, a RLO hot Jupiter either gets entire swallowed by the host star or leave behind a remnant planet with a thick atmosphere. Neither scenario would produce the observed sample of rocky hot Earths.
The rate of hydrodynamic mass loss (including photoevaporation) is strongly dependent on planet mass: while the H/He envelope of a $5\,M_\oplus$ planet can be easily dissipated over
100\,Myr, the timescale can be longer than a Hubble time for planets more massive than Neptune \citep[e.g.][]{Wang}. Hence, photoevaporation is unable to strip a hot Jupiter down to a hot Earth.

Thus, it seems possible or even probable that when we observe a hot Earth,
we are really looking at the exposed rocky core of a sub-Neptune.
If so, then the hot Earths are a sample of rocky cores that managed to
avoid runaway gas accretion and promotion into a giant
planet. With these premises, we can use the maximum observed mass of a hot Earth
as an indicator of the critical mass for initiating runaway gas
accretion. We note that the selection bias of radial-velocity mass measurements towards higher-mass planets works in our favor, since we are interested establishing
the maximum mass.

The critical mass for runaway gas accretion has been studied
for decades.
Assuming a radiative envelope, the critical mass has only a
logarithmic dependence on the local disk properties, and is often taken
to be $10\,M_\oplus$ \citep[e.g.][]{Rafikov}. More recently, \citet{Lee2016} advocated for the importance of envelope opacity and revised the threshold mass to be between 2 and $8\,M_\oplus$. Our uniform analysis of hot Earths seems to favor an upper limit of about $8\,M_\oplus$. This assumes that the planetary masses have not been significantly reduced by giant impact collisions, an issue we will visit in the next section.

\subsection{Mercury formation scenario}

Mercury has an iron mass fraction of about 70\%, which is significantly
larger than any of the other
planets in the solar system \citep{Hauck}.
Another explanation is a giant impact collision that removed much of the planet's
rocky mantle \citep{Benz+2008}.  \citet{Bonomo2019} also invoked a giant impact scenario to explain why Kepler-107c attained a much higher mean density than the similarly-sized innermost planet b.
Alternatively, \citet{Lewis} attributed the iron enhancement
of Mercury to the high temperatures that prevail near the Sun,
which favor iron condensation.
More recently, \citet{Wurm} called
attention to the process of photophoresis that separates grains of different thermal conductivity and creates a compositional gradient in the disk.

In all of these scenarios, one might expect hot Earths to be just as iron-enriched
as Mercury, or even more so. The orbital velocity of a hot
Earth is much larger, leading to more disruptive giant impact collisions.
The much hotter conditions of hot Earths should also make the effect of condensation temperature and photophoresis more pronounced. Our results suggest that although a few hot Earths do show
evidence for iron enhancement (K2-141b and K2-229b), a Mercury-like
composition is not the norm.
Could this mean that an Earth-like composition is universal for all rocky worlds \citep{Dressing}? Could it also mean that the conditions for removing
mantle material through giant impacts are relatively rare?

Enlarging the sample of hot Earths --- especially those with host stars
bright enough for mass measurements --- is a high priority for future work.
The ongoing NASA {\it Transiting Exoplanet Survey Satellite} ({\it TESS}) mission \citep{Ricker} will help in this regard. This nearly all-sky transit survey should allow the sample of hot Earths
to be doubled. More importantly, the host stars will be systematically brighter
than the typical stars in the {\it Kepler} and {\it K2} surveys.
{\it TESS} will also survey stars over a broader range of spectral types.
In combination with the suite of new Doppler spectrographs coming online, the larger and more diverse sample will hopefully improve our understanding of these extreme
worlds.

\begin{figure*}
\begin{center}
\includegraphics[width = 1.5\columnwidth]{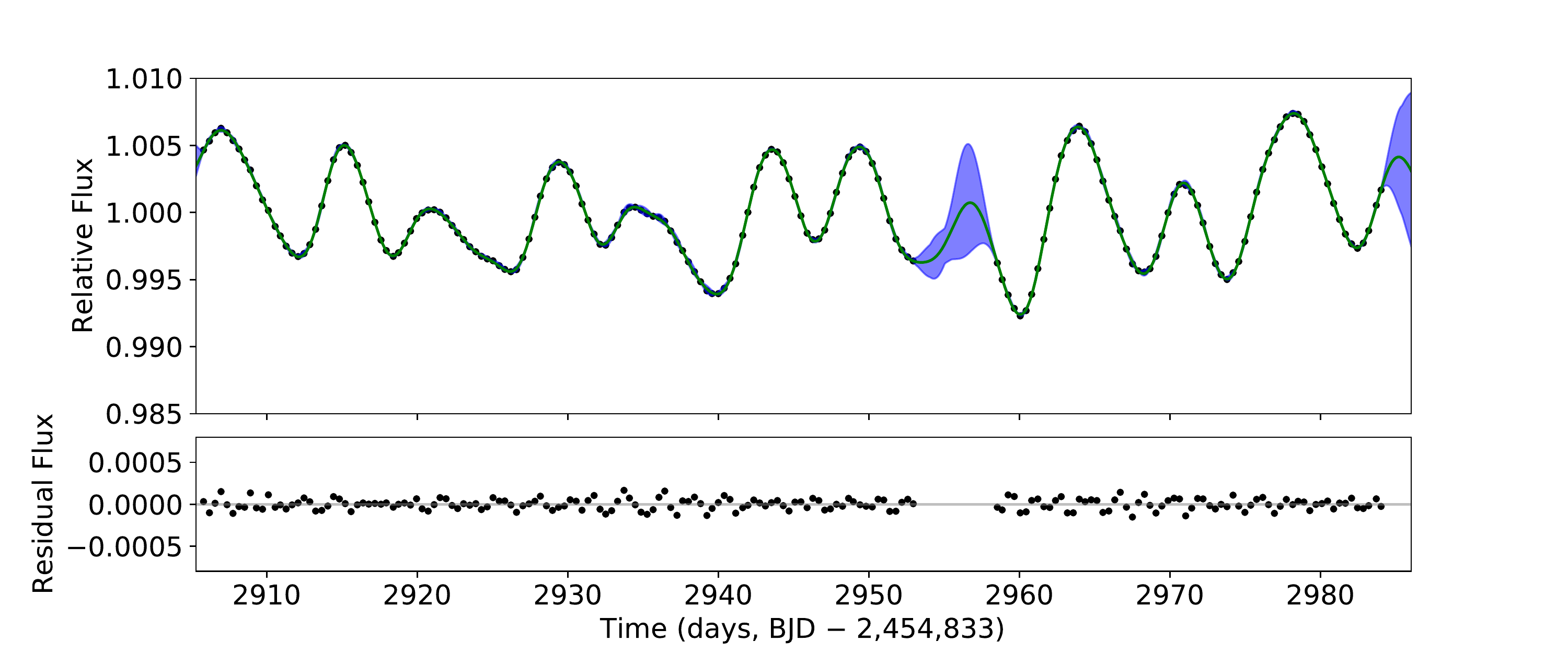}
\includegraphics[width = 1.5\columnwidth]{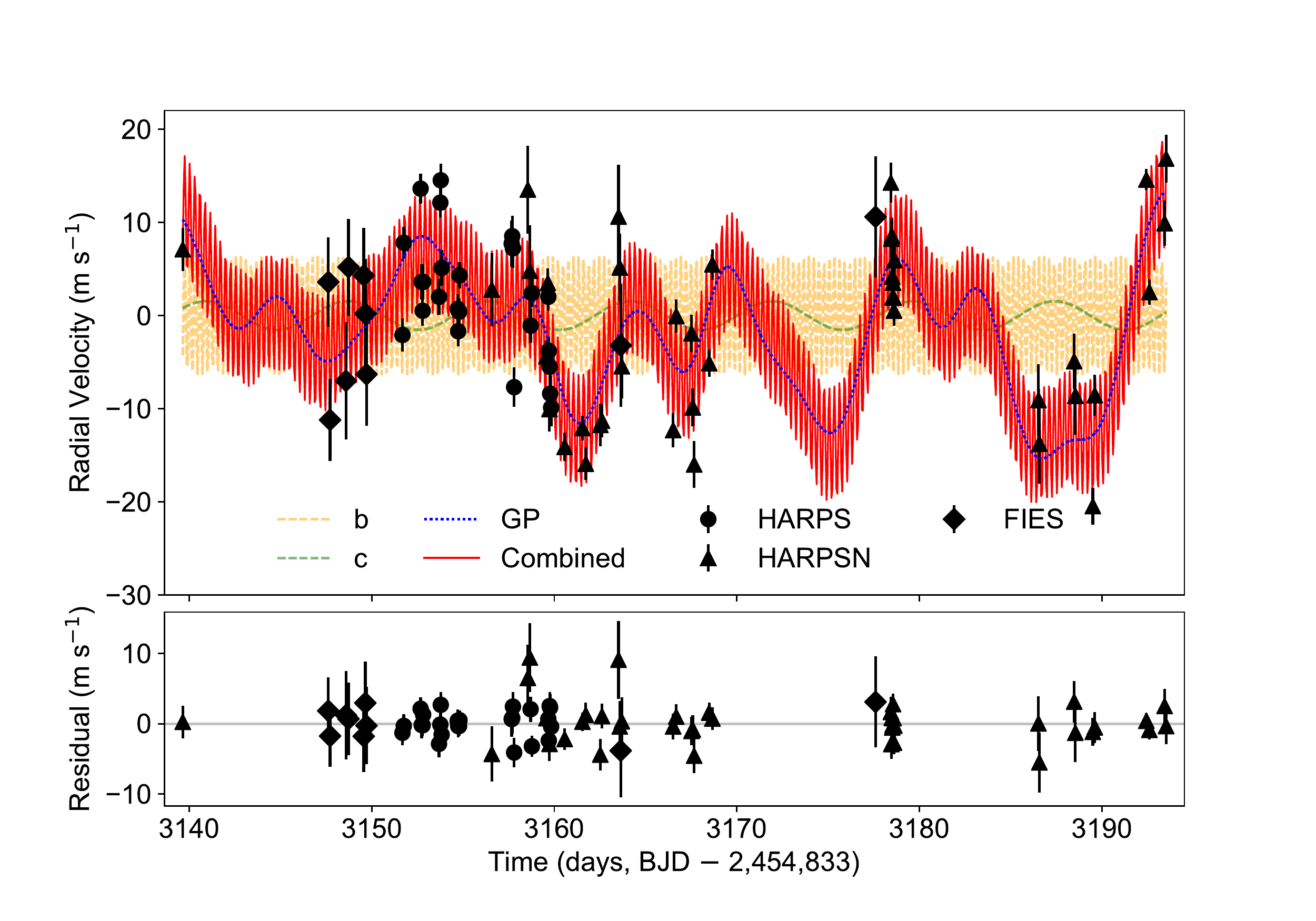}
\includegraphics[width = 1.5\columnwidth]{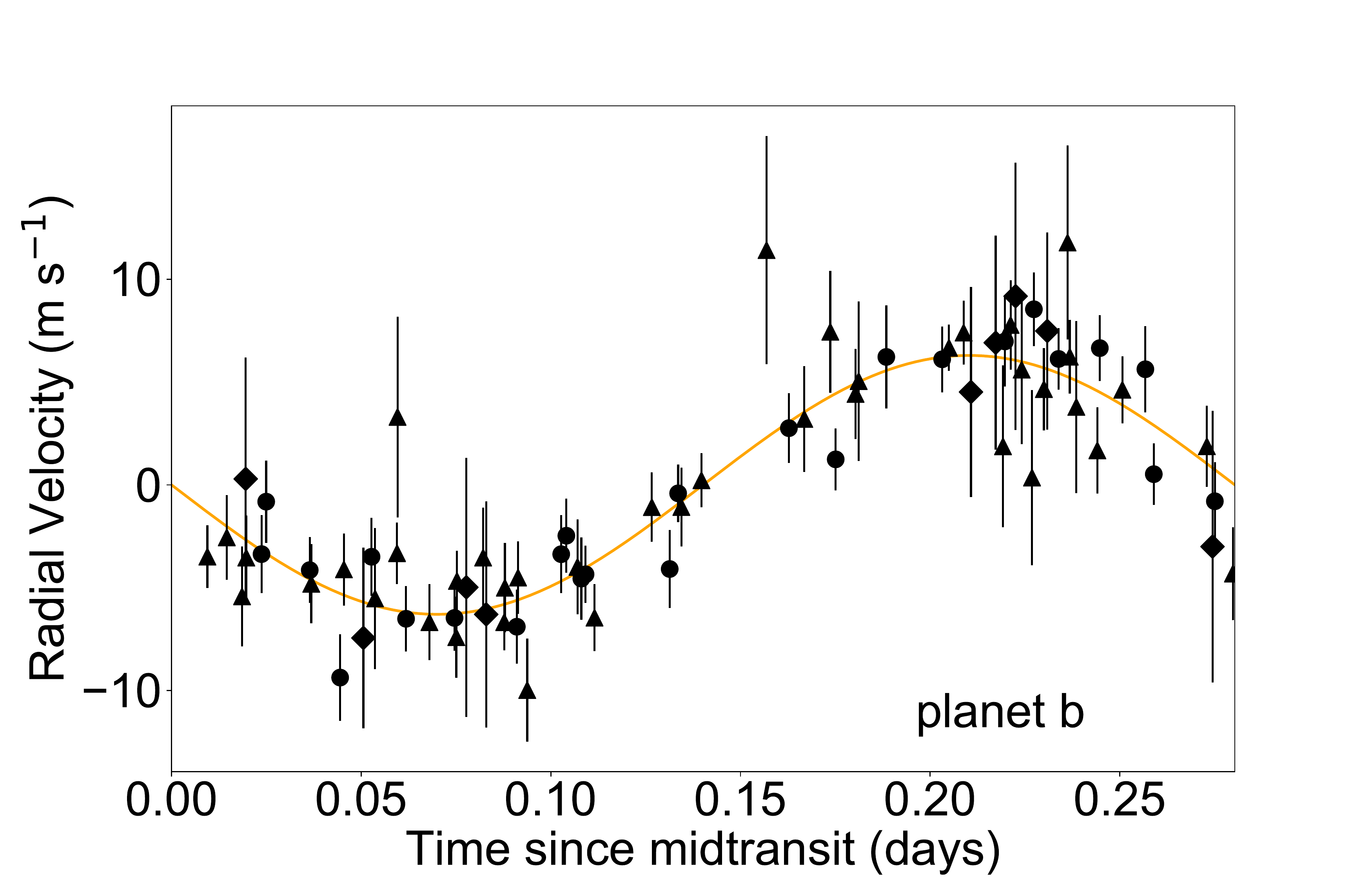}
\caption{Illustrative example of our Gaussian Process radial-velocity analysis,
for the K2-141 system.
{\it Top}: The out-of-transit flux variation is used to train the hyperparameters of the Gaussian Process model. {\it Middle}: Measured and calculated
radial velocities of planets b and c.
{\it Bottom}: Data and best-fitting model for planet b, after subtracting
the calculated radial-velocity variation due to planet c and stellar activity.}
\label{fig:fit}
\end{center}
\end{figure*}

\begin{figure*}
\begin{center}
\includegraphics[width = 1.9\columnwidth]{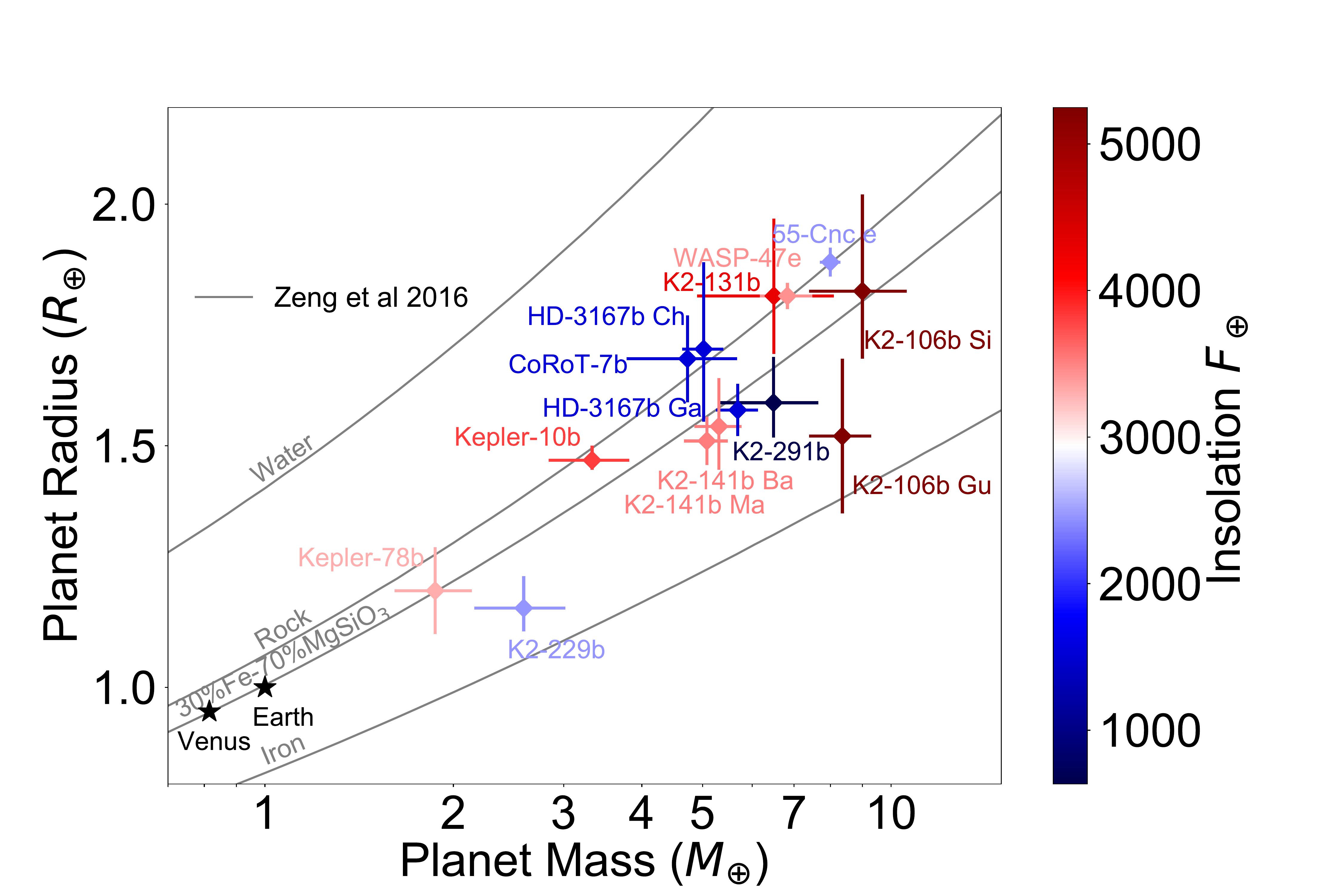}
\includegraphics[width = 1.9\columnwidth]{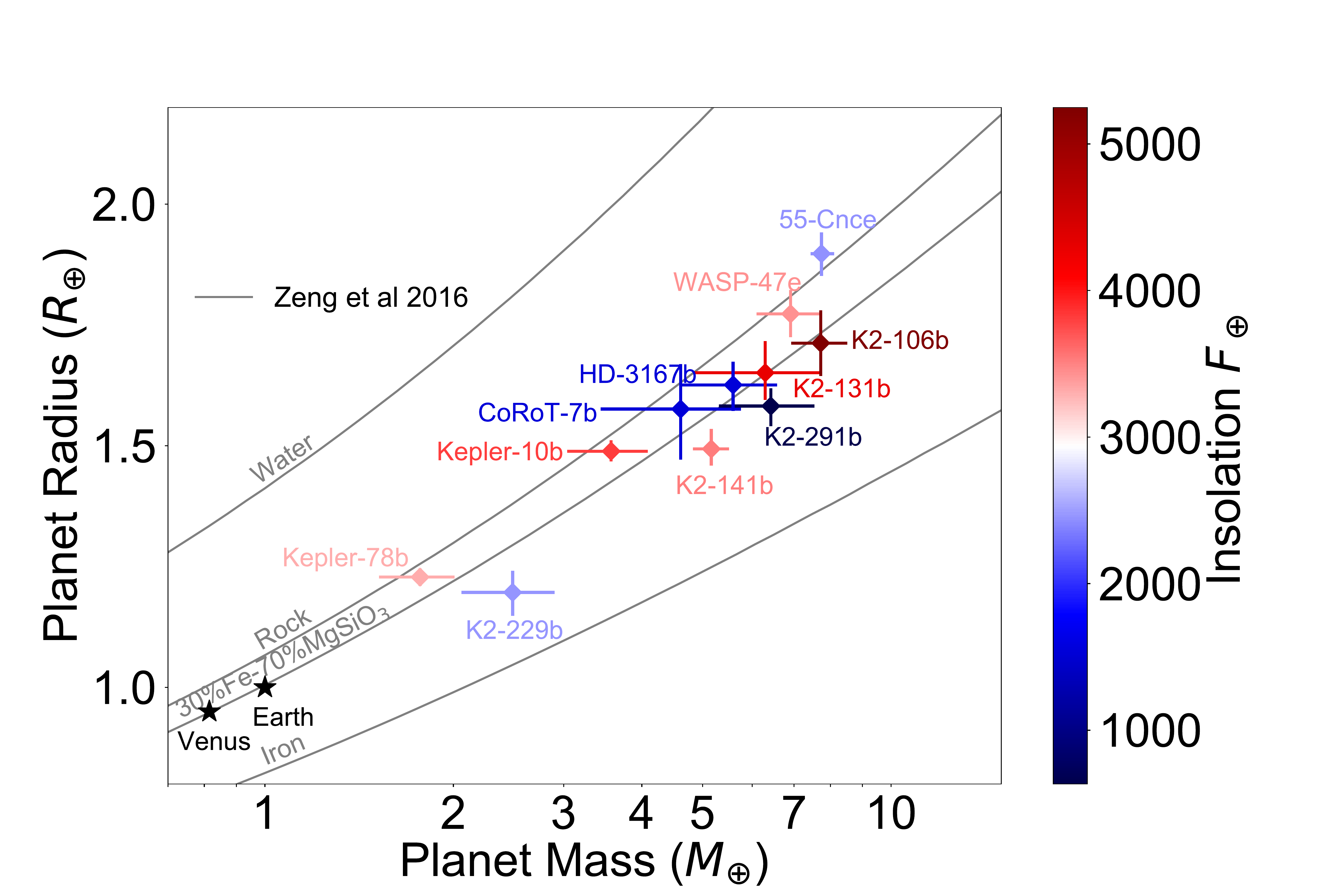}
\caption{Masses and radii of hot Earths in our sample
($R_{\rm p}/R)_\oplus <2$, $F/F_\oplus >650$, $\log g$ > 4.0)
and theoretical mass-radius curves for various compositions
from \citet{Zeng2016}.
The color of the symbols conveys the value of $F/F_\oplus$.
{\it Top}: Results drawn from the literature. For the cases
in which more than one report is available, the leading
initials of the first author are printed (see Table 2).
{\it Bottom}: Results from our analysis. In general,
the uncertainties are lower, and the dispersion in
the implied composition is smaller.}
\label{fig:fit}
\end{center}
\end{figure*}

\acknowledgements
 This work has made use of data from the European Space Agency (ESA) mission
{\it Gaia} (\url{https://www.cosmos.esa.int/gaia}), processed by the {\it Gaia}
Data Processing and Analysis Consortium (DPAC,
\url{https://www.cosmos.esa.int/web/gaia/dpac/consortium}). Funding for the DPAC
has been provided by national institutions, in particular the institutions
participating in the {\it Gaia} Multilateral Agreement.

\bibliography{main}

\begin{thebibliography}{}
\expandafter\ifx\csname natexlab\endcsname\relax\def\natexlab#1{#1}\fi
\providecommand{\url}[1]{\href{#1}{#1}}

\bibitem[{{Angelo} \& {Hu}(2017)}]{Angelo}
{Angelo}, I., \& {Hu}, R. 2017, \aj, 154, 232

\bibitem[{{Bailer-Jones} {et~al.}(2018){Bailer-Jones}, {Rybizki}, {Fouesneau},
  {Mantelet}, \& {Andrae}}]{2018AJ....156...58B}
{Bailer-Jones}, C.~A.~L., {Rybizki}, J., {Fouesneau}, M., {Mantelet}, G., \&
  {Andrae}, R. 2018, \aj, 156, 58

\bibitem[{{Barrag{\'a}n} {et~al.}(2018){Barrag{\'a}n}, {Gandolfi}, {Dai},
  {Livingston}, {Persson}, {Hirano}, {Narita}, {Csizmadia}, {Winn}, {Nespral},
  {Prieto-Arranz}, {Smith}, {Nowak}, {Albrecht}, {Antoniciello}, {Bo Justesen},
  {Cabrera}, {Cochran}, {Deeg}, {Eigmuller}, {Endl}, {Erikson}, {Fridlund},
  {Fukui}, {Grziwa}, {Guenther}, {Hatzes}, {Hidalgo}, {Johnson}, {Korth},
  {Palle}, {Patzold}, {Rauer}, {Tanaka}, \& {Van Eylen}}]{Barragan}
{Barrag{\'a}n}, O., {Gandolfi}, D., {Dai}, F., {et~al.} 2018, \aap, 612, A95

\bibitem[{{Becker} {et~al.}(2015){Becker}, {Vanderburg}, {Adams}, {Rappaport},
  \& {Schwengeler}}]{Becker}
{Becker}, J.~C., {Vanderburg}, A., {Adams}, F.~C., {Rappaport}, S.~A., \&
  {Schwengeler}, H.~M. 2015, \apj, 812, L18

\bibitem[{{Benz} {et~al.}(2008){Benz}, {Anic}, {Horner}, \&
  {Whitby}}]{Benz+2008}
{Benz}, W., {Anic}, A., {Horner}, J., \& {Whitby}, J.~A. 2008, {The Origin of
  Mercury}, ed. A.~{Balogh}, L.~{Ksanfomality}, \& R.~{von Steiger}, 7

\bibitem[{{Bonomo} {et~al.}(2019){Bonomo}, {Zeng}, {Damasso}, {Leinhardt},
  {Justesen}, {Lopez}, {Lund}, {Malavolta}, {Silva Aguirre}, \&
  {Buchhave}}]{Bonomo2019}
{Bonomo}, A.~S., {Zeng}, L., {Damasso}, M., {et~al.} 2019, Nature Astronomy, 3,
  416

\bibitem[{{Bourrier} {et~al.}(2018){Bourrier}, {Dumusque}, {Dorn}, {Henry},
  {Astudillo-Defru}, {Rey}, {Benneke}, {H{\'e}brard}, {Lovis}, {Demory},
  {Moutou}, \& {Ehrenreich}}]{Bourrier}
{Bourrier}, V., {Dumusque}, X., {Dorn}, C., {et~al.} 2018, \aap, 619, A1

\bibitem[{{Christiansen} {et~al.}(2017){Christiansen}, {Vanderburg}, {Burt},
  {Fulton}, {Batygin}, {Benneke}, {Brewer}, {Charbonneau}, {Ciardi}, {Collier
  Cameron}, {Coughlin}, {Crossfield}, {Dressing}, {Greene}, {Howard}, {Latham},
  {Molinari}, {Mortier}, {Mullally}, {Pepe}, {Rice}, {Sinukoff}, {Sozzetti},
  {Thompson}, {Udry}, {Vogt}, {Barman}, {Batalha}, {Bouchy}, {Buchhave},
  {Butler}, {Cosentino}, {Dupuy}, {Ehrenreich}, {Fiorenzano}, {Hansen},
  {Henning}, {Hirsch}, {Holden}, {Isaacson}, {Johnson}, {Knutson}, {Kosiarek},
  {L{\'o}pez-Morales}, {Lovis}, {Malavolta}, {Mayor}, {Micela}, {Motalebi},
  {Petigura}, {Phillips}, {Piotto}, {Rogers}, {Sasselov}, {Schlieder},
  {S{\'e}gransan}, {Watson}, \& {Weiss}}]{Christiansen2017}
{Christiansen}, J.~L., {Vanderburg}, A., {Burt}, J., {et~al.} 2017, \aj, 154,
  122

\bibitem[{{Dai} {et~al.}(2018){Dai}, {Masuda}, \& {Winn}}]{Dai2018}
{Dai}, F., {Masuda}, K., \& {Winn}, J.~N. 2018, \apjl, 864, L38

\bibitem[{{Dai} {et~al.}(2017){Dai}, {Winn}, {Gandolfi}, {Wang}, {Teske},
  {Burt}, {Albrecht}, {Barrag{\'a}n}, {Cochran}, {Endl}, {Fridlund}, {Hatzes},
  {Hirano}, {Hirsch}, {Johnson}, {Justesen}, {Livingston}, {Persson},
  {Prieto-Arranz}, {Vanderburg}, {Alonso}, {Antoniciello}, {Arriagada},
  {Butler}, {Cabrera}, {Crane}, {Cusano}, {Csizmadia}, {Deeg}, {Dieterich},
  {Eigm{\"u}ller}, {Erikson}, {Everett}, {Fukui}, {Grziwa}, {Guenther},
  {Henry}, {Howell}, {Johnson}, {Korth}, {Kuzuhara}, {Narita}, {Nespral},
  {Nowak}, {Palle}, {P{\"a}tzold}, {Rauer}, {Monta{\~n}{\'e}s Rodr{\'\i}guez},
  {Shectman}, {Smith}, {Thompson}, {Van Eylen}, {Williamson}, \&
  {Wittenmyer}}]{Dai2017}
{Dai}, F., {Winn}, J.~N., {Gandolfi}, D., {et~al.} 2017, \aj, 154, 226

\bibitem[{{Demory} {et~al.}(2016){Demory}, {Gillon}, {de Wit}, {Madhusudhan},
  {Bolmont}, {Heng}, {Kataria}, {Lewis}, {Hu}, {Krick}, {Stamenkovi{\'c}},
  {Benneke}, {Kane}, \& {Queloz}}]{Demory}
{Demory}, B.-O., {Gillon}, M., {de Wit}, J., {et~al.} 2016, \nat, 532, 207

\bibitem[{{Dorn} {et~al.}(2019){Dorn}, {Harrison}, {Bonsor}, \& {Hands}}]{Dorn}
{Dorn}, C., {Harrison}, J.~H.~D., {Bonsor}, A., \& {Hands}, T.~O. 2019, \mnras,
  484, 712

\bibitem[{{Dotter}(2016)}]{2016ApJS..222....8D}
{Dotter}, A. 2016, \apjs, 222, 8

\bibitem[{{Dressing} {et~al.}(2015){Dressing}, {Charbonneau}, {Dumusque},
  {Gettel}, {Pepe}, {Collier Cameron}, {Latham}, {Molinari}, {Udry}, {Affer},
  {Bonomo}, {Buchhave}, {Cosentino}, {Figueira}, {Fiorenzano}, {Harutyunyan},
  {Haywood}, {Johnson}, {Lopez-Morales}, {Lovis}, {Malavolta}, {Mayor},
  {Micela}, {Motalebi}, {Nascimbeni}, {Phillips}, {Piotto}, {Pollacco},
  {Queloz}, {Rice}, {Sasselov}, {S{\'e}gransan}, {Sozzetti}, {Szentgyorgyi}, \&
  {Watson}}]{Dressing}
{Dressing}, C.~D., {Charbonneau}, D., {Dumusque}, X., {et~al.} 2015, \apj, 800,
  135

\bibitem[{{Dumusque} {et~al.}(2014){Dumusque}, {Bonomo}, {Haywood},
  {Malavolta}, {S{\'e}gransan}, {Buchhave}, {Collier Cameron}, {Latham},
  {Molinari}, {Pepe}, {Udry}, {Charbonneau}, {Cosentino}, {Dressing},
  {Figueira}, {Fiorenzano}, {Gettel}, {Harutyunyan}, {Horne}, {Lopez-Morales},
  {Lovis}, {Mayor}, {Micela}, {Motalebi}, {Nascimbeni}, {Phillips}, {Piotto},
  {Pollacco}, {Queloz}, {Rice}, {Sasselov}, {Sozzetti}, {Szentgyorgyi}, \&
  {Watson}}]{Dumusque2014}
{Dumusque}, X., {Bonomo}, A.~S., {Haywood}, R.~D., {et~al.} 2014, \apj, 789,
  154

\bibitem[{{Eastman} {et~al.}(2013){Eastman}, {Gaudi}, \& {Agol}}]{Eastman2013}
{Eastman}, J., {Gaudi}, B.~S., \& {Agol}, E. 2013, \pasp, 125, 83

\bibitem[{{Foreman-Mackey} {et~al.}(2013){Foreman-Mackey}, {Hogg}, {Lang}, \&
  {Goodman}}]{emcee}
{Foreman-Mackey}, D., {Hogg}, D.~W., {Lang}, D., \& {Goodman}, J. 2013, \pasp,
  125, 306

\bibitem[{{Fulton} \& {Petigura}(2018)}]{Fulton2018}
{Fulton}, B.~J., \& {Petigura}, E.~A. 2018, \aj, 156, 264

\bibitem[{{Fulton} {et~al.}(2017){Fulton}, {Petigura}, {Howard}, {Isaacson},
  {Marcy}, {Cargile}, {Hebb}, {Weiss}, {Johnson}, {Morton}, {Sinukoff},
  {Crossfield}, \& {Hirsch}}]{Fulton}
{Fulton}, B.~J., {Petigura}, E.~A., {Howard}, A.~W., {et~al.} 2017, \aj, 154,
  109

\bibitem[{{Gaia Collaboration} {et~al.}(2018){Gaia Collaboration}, {Brown},
  {Vallenari}, {Prusti}, {de Bruijne}, {Babusiaux}, {Bailer-Jones}, {Biermann},
  {Evans}, {Eyer}, \& et~al.}]{2018A&A...616A...1G}
{Gaia Collaboration}, {Brown}, A.~G.~A., {Vallenari}, A., {et~al.} 2018, \aap,
  616, A1

\bibitem[{{Gandolfi} {et~al.}(2017){Gandolfi}, {Barrag{\'a}n}, {Hatzes},
  {Fridlund}, {Fossati}, {Donati}, {Johnson}, {Nowak}, {Prieto-Arranz},
  {Albrecht}, {Dai}, {Deeg}, {Endl}, {Grziwa}, {Hjorth}, {Korth}, {Nespral},
  {Saario}, {Smith}, {Antoniciello}, {Alarcon}, {Bedell}, {Blay}, {Brems},
  {Cabrera}, {Csizmadia}, {Cusano}, {Cochran}, {Eigm{\"u}ller}, {Erikson},
  {Gonz{\'a}lez Hern{\'a}ndez}, {Guenther}, {Hirano}, {Su{\'a}rez
  Mascare{\~n}o}, {Narita}, {Palle}, {Parviainen}, {P{\"a}tzold}, {Persson},
  {Rauer}, {Saviane}, {Schmidtobreick}, {Van Eylen}, {Winn}, \&
  {Zakhozhay}}]{Gandolfi}
{Gandolfi}, D., {Barrag{\'a}n}, O., {Hatzes}, A.~P., {et~al.} 2017, \aj, 154,
  123

\bibitem[{{Ginzburg} \& {Sari}(2017)}]{Ginzburg}
{Ginzburg}, S., \& {Sari}, R. 2017, \mnras, 469, 278

\bibitem[{{Green} {et~al.}(2018){Green}, {Schlafly}, {Finkbeiner}, {Rix},
  {Martin}, {Burgett}, {Draper}, {Flewelling}, {Hodapp}, {Kaiser}, {Kudritzki},
  {Magnier}, {Metcalfe}, {Tonry}, {Wainscoat}, \&
  {Waters}}]{2018MNRAS.478..651G}
{Green}, G.~M., {Schlafly}, E.~F., {Finkbeiner}, D., {et~al.} 2018, \mnras,
  478, 651

\bibitem[{{Guenther} {et~al.}(2017){Guenther}, {Barrag{\'a}n}, {Dai},
  {Gandolfi}, {Hirano}, {Fridlund}, {Fossati}, {Chau}, {Helled}, {Korth},
  {Prieto-Arranz}, {Nespral}, {Antoniciello}, {Deeg}, {Hjorth}, {Grziwa},
  {Albrecht}, {Hatzes}, {Rauer}, {Csizmadia}, {Smith}, {Cabrera}, {Narita},
  {Arriagada}, {Burt}, {Butler}, {Cochran}, {Crane}, {Eigm{\"u}ller},
  {Erikson}, {Johnson}, {Kiilerich}, {Kubyshkina}, {Palle}, {Persson},
  {P{\"a}tzold}, {Sabotta}, {Sato}, {Shectman}, {Teske}, {Thompson}, {Van
  Eylen}, {Nowak}, {Vanderburg}, {Winn}, \& {Wittenmyer}}]{Guenther}
{Guenther}, E.~W., {Barrag{\'a}n}, O., {Dai}, F., {et~al.} 2017, \aap, 608, A93

\bibitem[{Hauck~II {et~al.}(2013)Hauck~II, Margot, Solomon, Phillips, Johnson,
  Lemoine, Mazarico, McCoy, Padovan, Peale, Perry, Smith, \& Zuber}]{Hauck}
Hauck~II, S.~A., Margot, J.-L., Solomon, S.~C., {et~al.} 2013, Journal of
  Geophysical Research: Planets, 118, 1204.
\newblock
  \url{https://agupubs.onlinelibrary.wiley.com/doi/abs/10.1002/jgre.20091}

\bibitem[{{Haywood} {et~al.}(2014){Haywood}, {Collier Cameron}, {Queloz},
  {Barros}, {Deleuil}, {Fares}, {Gillon}, {Lanza}, {Lovis}, {Moutou}, {Pepe},
  {Pollacco}, {Santerne}, {S{\'e}gransan}, \& {Unruh}}]{Haywood}
{Haywood}, R.~D., {Collier Cameron}, A., {Queloz}, D., {et~al.} 2014, \mnras,
  443, 2517

\bibitem[{{Howard} {et~al.}(2013){Howard}, {Sanchis-Ojeda}, {Marcy}, {Johnson},
  {Winn}, {Isaacson}, {Fischer}, {Fulton}, {Sinukoff}, \&
  {Fortney}}]{Howard2013}
{Howard}, A.~W., {Sanchis-Ojeda}, R., {Marcy}, G.~W., {et~al.} 2013, \nat, 503,
  381

\bibitem[{{Isella} {et~al.}(2006){Isella}, {Testi}, \& {Natta}}]{Isella}
{Isella}, A., {Testi}, L., \& {Natta}, A. 2006, \aap, 451, 951

\bibitem[{{Kosiarek} {et~al.}(2019){Kosiarek}, {Blunt}, {L{\'o}pez-Morales},
  {Crossfield}, {Sinukoff}, {Petigura}, {Gonzales}, {Poretti}, {Malavolta},
  {Howard}, {Isaacson}, {Haywood}, {Ciardi}, {Bristow}, {Collier Cameron},
  {Charbonneau}, {Dressing}, {Figueira}, {Fulton}, {Hardee}, {Hirsch},
  {Latham}, {Mortier}, {Nava}, {Schlieder}, {Vanderburg}, {Weiss}, {Bonomo},
  {Bouchy}, {Buchhave}, {Coffinet}, {Damasso}, {Dumusque}, {Lovis}, {Mayor},
  {Micela}, {Molinari}, {Pepe}, {Phillips}, {Piotto}, {Rice}, {Sasselov},
  {S{\'e}gransan}, {Sozzetti}, {Udry}, \& {Watson}}]{Kosiarek}
{Kosiarek}, M.~R., {Blunt}, S., {L{\'o}pez-Morales}, M., {et~al.} 2019, \aj,
  157, 116

\bibitem[{{Kreidberg}(2015)}]{Kreidberg2015}
{Kreidberg}, L. 2015, \pasp, 127, 1161

\bibitem[{{Lammer} {et~al.}(2009){Lammer}, {Odert}, {Leitzinger},
  {Khodachenko}, {Panchenko}, {Kulikov}, {Zhang}, {Lichtenegger}, {Erkaev},
  {Wuchterl}, {Micela}, {Penz}, {Biernat}, {Weingrill}, {Steller}, {Ottacher},
  {Hasiba}, \& {Hanslmeier}}]{Lammer2009}
{Lammer}, H., {Odert}, P., {Leitzinger}, M., {et~al.} 2009, \aap, 506, 399

\bibitem[{{Lee} \& {Chiang}(2016)}]{Lee2016}
{Lee}, E.~J., \& {Chiang}, E. 2016, \apj, 817, 90

\bibitem[{{Lee} \& {Chiang}(2017)}]{Lee}
---. 2017, \apj, 842, 40

\bibitem[{{L{\'e}ger} {et~al.}(2009){L{\'e}ger}, {Rouan}, {Schneider}, {Barge},
  {Fridlund}, {Samuel}, {Ollivier}, {Guenther}, {Deleuil}, {Deeg}, {Auvergne},
  {Alonso}, {Aigrain}, {Alapini}, {Almenara}, {Baglin}, {Barbieri}, {Bruntt},
  {Bord{\'e}}, {Bouchy}, {Cabrera}, {Catala}, {Carone}, {Carpano}, {Csizmadia},
  {Dvorak}, {Erikson}, {Ferraz-Mello}, {Foing}, {Fressin}, {Gandolfi},
  {Gillon}, {Gondoin}, {Grasset}, {Guillot}, {Hatzes}, {H{\'e}brard}, {Jorda},
  {Lammer}, {Llebaria}, {Loeillet}, {Mayor}, {Mazeh}, {Moutou}, {P{\"a}tzold},
  {Pont}, {Queloz}, {Rauer}, {Renner}, {Samadi}, {Shporer}, {Sotin}, {Tingley},
  {Wuchterl}, {Adda}, {Agogu}, {Appourchaux}, {Ballans}, {Baron}, {Beaufort},
  {Bellenger}, {Berlin}, {Bernardi}, {Blouin}, {Baudin}, {Bodin}, {Boisnard},
  {Boit}, {Bonneau}, {Borzeix}, {Briet}, {Buey}, {Butler}, {Cailleau},
  {Cautain}, {Chabaud}, {Chaintreuil}, {Chiavassa}, {Costes}, {Cuna Parrho},
  {de Oliveira Fialho}, {Decaudin}, {Defise}, {Djalal}, {Epstein}, {Exil},
  {Faur{\'e}}, {Fenouillet}, {Gaboriaud}, {Gallic}, {Gamet}, {Gavalda},
  {Grolleau}, {Gruneisen}, {Gueguen}, {Guis}, {Guivarc'h}, {Guterman},
  {Hallouard}, {Hasiba}, {Heuripeau}, {Huntzinger}, {Hustaix}, {Imad},
  {Imbert}, {Johlander}, {Jouret}, {Journoud}, {Karioty}, {Kerjean},
  {Lafaille}, {Lafond}, {Lam-Trong}, {Landiech}, {Lapeyrere}, {Larqu{\'e}},
  {Laudet}, {Lautier}, {Lecann}, {Lefevre}, {Leruyet}, {Levacher}, {Magnan},
  {Mazy}, {Mertens}, {Mesnager}, {Meunier}, {Michel}, {Monjoin}, {Naudet},
  {Nguyen-Kim}, {Orcesi}, {Ottacher}, {Perez}, {Peter}, {Plasson}, {Plesseria},
  {Pontet}, {Pradines}, {Quentin}, {Reynaud}, {Rolland}, {Rollenhagen},
  {Romagnan}, {Russ}, {Schmidt}, {Schwartz}, {Sebbag}, {Sedes}, {Smit},
  {Steller}, {Sunter}, {Surace}, {Tello}, {Tiph{\`e}ne}, {Toulouse}, {Ulmer},
  {Vandermarcq}, {Vergnault}, {Vuillemin}, \& {Zanatta}}]{Leger}
{L{\'e}ger}, A., {Rouan}, D., {Schneider}, J., {et~al.} 2009, \aap, 506, 287

\bibitem[{Lewis(1972)}]{Lewis}
Lewis, J.~S. 1972, Earth and Planetary Science Letters, 15, 286 .
\newblock
  \url{http://www.sciencedirect.com/science/article/pii/0012821X72901744}

\bibitem[{{Lopez}(2017)}]{Lopez2017}
{Lopez}, E.~D. 2017, \mnras, 472, 245

\bibitem[{{Lundkvist} {et~al.}(2016){Lundkvist}, {Kjeldsen}, {Albrecht},
  {Davies}, {Basu}, {Huber}, {Justesen}, {Karoff}, {Silva Aguirre}, {van
  Eylen}, {Vang}, {Arentoft}, {Barclay}, {Bedding}, {Campante}, {Chaplin},
  {Christensen-Dalsgaard}, {Elsworth}, {Gilliland}, {Handberg}, {Hekker},
  {Kawaler}, {Lund}, {Metcalfe}, {Miglio}, {Rowe}, {Stello}, {Tingley}, \&
  {White}}]{Lundkvist}
{Lundkvist}, M.~S., {Kjeldsen}, H., {Albrecht}, S., {et~al.} 2016, Nature
  Communications, 7, 11201

\bibitem[{{Malavolta} {et~al.}(2018){Malavolta}, {Mayo}, {Louden}, {Rajpaul},
  {Bonomo}, {Buchhave}, {Kreidberg}, {Kristiansen}, {Lopez-Morales}, {Mortier},
  {Vand erburg}, {Coffinet}, {Ehrenreich}, {Lovis}, {Bouchy}, {Charbonneau},
  {Ciardi}, {Collier Cameron}, {Cosentino}, {Crossfield}, {Damasso},
  {Dressing}, {Dumusque}, {Everett}, {Figueira}, {Fiorenzano}, {Gonzales},
  {Haywood}, {Harutyunyan}, {Hirsch}, {Howell}, {Johnson}, {Latham}, {Lopez},
  {Mayor}, {Micela}, {Molinari}, {Nascimbeni}, {Pepe}, {Phillips}, {Piotto},
  {Rice}, {Sasselov}, {S{\'e}gransan}, {Sozzetti}, {Udry}, \&
  {Watson}}]{Malavolta2018}
{Malavolta}, L., {Mayo}, A.~W., {Louden}, T., {et~al.} 2018, \aj, 155, 107

\bibitem[{{Morton}(2015)}]{2015ascl.soft03010M}
{Morton}, T.~D. 2015, {isochrones: Stellar model grid package}, Astrophysics
  Source Code Library, , , ascl:1503.010

\bibitem[{{Niraula} {et~al.}(2017){Niraula}, {Redfield}, {Dai}, {Barrag{\'a}n},
  {Gandolfi}, {Cauley}, {Hirano}, {Korth}, {Smith}, {Prieto-Arranz}, {Grziwa},
  {Fridlund}, {Persson}, {Justesen}, {Winn}, {Albrecht}, {Cochran},
  {Csizmadia}, {Duvvuri}, {Endl}, {Hatzes}, {Livingston}, {Narita}, {Nespral},
  {Nowak}, {P{\"a}tzold}, {Palle}, \& {Van Eylen}}]{Niraula2017}
{Niraula}, P., {Redfield}, S., {Dai}, F., {et~al.} 2017, \aj, 154, 266

\bibitem[{{Ofir} {et~al.}(2018){Ofir}, {Xie}, {Jiang}, {Sari}, \&
  {Aharonson}}]{Ofir}
{Ofir}, A., {Xie}, J.-W., {Jiang}, C.-F., {Sari}, R., \& {Aharonson}, O. 2018,
  \apjs, 234, 9

\bibitem[{{Owen} \& {Wu}(2017)}]{Owen}
{Owen}, J.~E., \& {Wu}, Y. 2017, \apj, 847, 29

\bibitem[{{Paxton} {et~al.}(2011){Paxton}, {Bildsten}, {Dotter}, {Herwig},
  {Lesaffre}, \& {Timmes}}]{Paxton2011}
{Paxton}, B., {Bildsten}, L., {Dotter}, A., {et~al.} 2011, \apjs, 192, 3

\bibitem[{{Pepe} {et~al.}(2013){Pepe}, {Cameron}, {Latham}, {Molinari}, {Udry},
  {Bonomo}, {Buchhave}, {Charbonneau}, {Cosentino}, {Dressing}, {Dumusque},
  {Figueira}, {Fiorenzano}, {Gettel}, {Harutyunyan}, {Haywood}, {Horne},
  {Lopez-Morales}, {Lovis}, {Malavolta}, {Mayor}, {Micela}, {Motalebi},
  {Nascimbeni}, {Phillips}, {Piotto}, {Pollacco}, {Queloz}, {Rice}, {Sasselov},
  {S{\'e}gransan}, {Sozzetti}, {Szentgyorgyi}, \& {Watson}}]{Pepe}
{Pepe}, F., {Cameron}, A.~C., {Latham}, D.~W., {et~al.} 2013, \nat, 503, 377

\bibitem[{{Petigura} {et~al.}(2018){Petigura}, {Marcy}, {Winn}, {Weiss},
  {Fulton}, {Howard}, {Sinukoff}, {Isaacson}, {Morton}, \&
  {Johnson}}]{Petigura}
{Petigura}, E.~A., {Marcy}, G.~W., {Winn}, J.~N., {et~al.} 2018, \aj, 155, 89

\bibitem[{{Prieto-Arranz} {et~al.}(2018){Prieto-Arranz}, {Palle}, {Gandolfi},
  {Barrag{\'a}n}, {Guenther}, {Dai}, {Fridlund}, {Hirano}, {Livingston},
  {Luque}, {Niraula}, {Persson}, {Redfield}, {Albrecht}, {Alonso},
  {Antoniciello}, {Cabrera}, {Cochran}, {Csizmadia}, {Deeg}, {Eigm{\"u}ller},
  {Endl}, {Erikson}, {Everett}, {Fukui}, {Grziwa}, {Hatzes}, {Hidalgo},
  {Hjorth}, {Korth}, {Lorenzo-Oliveira}, {Murgas}, {Narita}, {Nespral},
  {Nowak}, {P{\"a}tzold}, {Monta{\~n}ez Rodr{\'\i}guez}, {Rauer}, {Ribas},
  {Smith}, {Trifonov}, {Van Eylen}, \& {Winn}}]{Prieto}
{Prieto-Arranz}, J., {Palle}, E., {Gandolfi}, D., {et~al.} 2018, \aap, 618,
  A116

\bibitem[{{Rafikov}(2006)}]{Rafikov}
{Rafikov}, R.~R. 2006, \apj, 648, 666

\bibitem[{{Rice} {et~al.}(2019){Rice}, {Malavolta}, {Mayo}, {Mortier},
  {Buchhave}, {Affer}, {Vanderburg}, {Lopez-Morales}, {Poretti}, {Zeng},
  {Cameron}, {Damasso}, {Coffinet}, {Latham}, {Bonomo}, {Bouchy},
  {Charbonneau}, {Dumusque}, {Figueira}, {Martinez Fiorenzano}, {Haywood},
  {Johnson}, {Lopez}, {Lovis}, {Mayor}, {Micela}, {Molinari}, {Nascimbeni},
  {Nava}, {Pepe}, {Phillips}, {Piotto}, {Sasselov}, {S{\'e}gransan},
  {Sozzetti}, {Udry}, \& {Watson}}]{Rice}
{Rice}, K., {Malavolta}, L., {Mayo}, A., {et~al.} 2019, \mnras, 484, 3731

\bibitem[{{Ricker} {et~al.}(2014){Ricker}, {Winn}, {Vanderspek}, {Latham},
  {Bakos}, {Bean}, {Berta-Thompson}, {Brown}, {Buchhave}, {Butler}, {Butler},
  {Chaplin}, {Charbonneau}, {Christensen-Dalsgaard}, {Clampin}, {Deming},
  {Doty}, {De Lee}, {Dressing}, {Dunham}, {Endl}, {Fressin}, {Ge}, {Henning},
  {Holman}, {Howard}, {Ida}, {Jenkins}, {Jernigan}, {Johnson}, {Kaltenegger},
  {Kawai}, {Kjeldsen}, {Laughlin}, {Levine}, {Lin}, {Lissauer}, {MacQueen},
  {Marcy}, {McCullough}, {Morton}, {Narita}, {Paegert}, {Palle}, {Pepe},
  {Pepper}, {Quirrenbach}, {Rinehart}, {Sasselov}, {Sato}, {Seager},
  {Sozzetti}, {Stassun}, {Sullivan}, {Szentgyorgyi}, {Torres}, {Udry}, \&
  {Villasenor}}]{Ricker}
{Ricker}, G.~R., {Winn}, J.~N., {Vanderspek}, R., {et~al.} 2014, in Society of
  Photo-Optical Instrumentation Engineers (SPIE) Conference Series, Vol. 9143,
  Space Telescopes and Instrumentation 2014: Optical, Infrared, and Millimeter
  Wave, 914320

\bibitem[{{Rogers}(2015)}]{Rogers}
{Rogers}, L.~A. 2015, \apj, 801, 41

\bibitem[{{Sanchis-Ojeda} {et~al.}(2014){Sanchis-Ojeda}, {Rappaport}, {Winn},
  {Kotson}, {Levine}, \& {El Mellah}}]{Sanchis}
{Sanchis-Ojeda}, R., {Rappaport}, S., {Winn}, J.~N., {et~al.} 2014, \apj, 787,
  47

\bibitem[{{Sanchis-Ojeda} {et~al.}(2013){Sanchis-Ojeda}, {Rappaport}, {Winn},
  {Levine}, {Kotson}, {Latham}, \& {Buchhave}}]{Sanchis2013}
---. 2013, \apj, 774, 54

\bibitem[{{Santerne} {et~al.}(2018){Santerne}, {Brugger}, {Armstrong},
  {Adibekyan}, {Lillo-Box}, {Gosselin}, {Aguichine}, {Almenara}, {Barrado},
  {Barros}, {Bayliss}, {Boisse}, {Bonomo}, {Bouchy}, {Brown}, {Deleuil},
  {Delgado Mena}, {Demangeon}, {D{\'\i}az}, {Doyle}, {Dumusque}, {Faedi},
  {Faria}, {Figueira}, {Foxell}, {Giles}, {H{\'e}brard}, {Hojjatpanah},
  {Hobson}, {Jackman}, {King}, {Kirk}, {Lam}, {Ligi}, {Lovis}, {Louden},
  {McCormac}, {Mousis}, {Neal}, {Osborn}, {Pepe}, {Pollacco}, {Santos},
  {Sousa}, {Udry}, \& {Vigan}}]{Santerne2018}
{Santerne}, A., {Brugger}, B., {Armstrong}, D.~J., {et~al.} 2018, Nature
  Astronomy, 2, 393

\bibitem[{{Sinukoff} {et~al.}(2017){Sinukoff}, {Howard}, {Petigura}, {Fulton},
  {Crossfield}, {Isaacson}, {Gonzales}, {Crepp}, {Brewer}, {Hirsch}, {Weiss},
  {Ciardi}, {Schlieder}, {Benneke}, {Christiansen}, {Dressing}, {Hansen},
  {Knutson}, {Kosiarek}, {Livingston}, {Greene}, {Rogers}, \&
  {L{\'e}pine}}]{Sinukoff2017}
{Sinukoff}, E., {Howard}, A.~W., {Petigura}, E.~A., {et~al.} 2017, \aj, 153,
  271

\bibitem[{{Skrutskie} {et~al.}(2006){Skrutskie}, {Cutri}, {Stiening},
  {Weinberg}, {Schneider}, {Carpenter}, {Beichman}, {Capps}, {Chester},
  {Elias}, {Huchra}, {Liebert}, {Lonsdale}, {Monet}, {Price}, {Seitzer},
  {Jarrett}, {Kirkpatrick}, {Gizis}, {Howard}, {Evans}, {Fowler}, {Fullmer},
  {Hurt}, {Light}, {Kopan}, {Marsh}, {McCallon}, {Tam}, {Van Dyk}, \&
  {Wheelock}}]{2006AJ....131.1163S}
{Skrutskie}, M.~F., {Cutri}, R.~M., {Stiening}, R., {et~al.} 2006, \aj, 131,
  1163

\bibitem[{{Southworth}(2008)}]{Southworth2008}
{Southworth}, J. 2008, \mnras, 386, 1644

\bibitem[{{Southworth}(2009)}]{Southworth2009}
---. 2009, \mnras, 394, 272

\bibitem[{{Teske} {et~al.}(2018){Teske}, {Wang}, {Wolfgang}, {Dai}, {Shectman},
  {Butler}, {Crane}, \& {Thompson}}]{Teske}
{Teske}, J.~K., {Wang}, S., {Wolfgang}, A., {et~al.} 2018, \aj, 155, 148

\bibitem[{{Torres} {et~al.}(2012){Torres}, {Fischer}, {Sozzetti}, {Buchhave},
  {Winn}, {Holman}, \& {Carter}}]{Torres2012}
{Torres}, G., {Fischer}, D.~A., {Sozzetti}, A., {et~al.} 2012, \apj, 757, 161

\bibitem[{{Valsecchi} {et~al.}(2015){Valsecchi}, {Rappaport}, {Rasio},
  {Marchant}, \& {Rogers}}]{Valsecchi}
{Valsecchi}, F., {Rappaport}, S., {Rasio}, F.~A., {Marchant}, P., \& {Rogers},
  L.~A. 2015, \apj, 813, 101

\bibitem[{{Vanderburg} {et~al.}(2017){Vanderburg}, {Becker}, {Buchhave},
  {Mortier}, {Lopez}, {Malavolta}, {Haywood}, {Latham}, {Charbonneau},
  {L{\'o}pez-Morales}, {Adams}, {Bonomo}, {Bouchy}, {Collier Cameron},
  {Cosentino}, {Di Fabrizio}, {Dumusque}, {Fiorenzano}, {Harutyunyan},
  {Johnson}, {Lorenzi}, {Lovis}, {Mayor}, {Micela}, {Molinari}, {Pedani},
  {Pepe}, {Piotto}, {Phillips}, {Rice}, {Sasselov}, {S{\'e}gransan},
  {Sozzetti}, {Udry}, \& {Watson}}]{Vanderburg2017}
{Vanderburg}, A., {Becker}, J.~C., {Buchhave}, L.~A., {et~al.} 2017, \aj, 154,
  237

\bibitem[{{Wang} \& {Dai}(2017)}]{Wang}
{Wang}, L., \& {Dai}, F. 2017, ArXiv e-prints, arXiv:1710.03826

\bibitem[{{West} {et~al.}(2018){West}, {Gillen}, {Bayliss}, {Burleigh},
  {Delrez}, {G{\"u}nther}, {Hodgkin}, {Jackman}, {Jenkins}, {King}, {McCormac},
  {Nielsen}, {Raynard}, {Smith}, {Soto}, {Turner}, {Wheatley}, {Almleaky},
  {Armstrong}, {Belardi}, {Bouchy}, {Briegal}, {Burdanov}, {Cabrera},
  {Casewel}, {Chaushev}, {Chazelas}, {Chote}, {Cooke}, {Csizmadia}, {Ducrot},
  {Eigm{\"u}ller}, {Erikson}, {Foxell}, {G{\"a}nsicke}, {Gillon}, {Goad},
  {Jehin}, {Lambert}, {Longstaff}, {Louden}, {Moyano}, {Murray}, {Pollacco},
  {Queloz}, {Rauer}, {Sohy}, {Thompson}, {Udry}, {Walker}, \& {Watson}}]{West}
{West}, R.~G., {Gillen}, E., {Bayliss}, D., {et~al.} 2018, arXiv e-prints,
  arXiv:1809.00678

\bibitem[{{Winn} {et~al.}(2018){Winn}, {Sanchis-Ojeda}, \&
  {Rappaport}}]{Winn2018}
{Winn}, J.~N., {Sanchis-Ojeda}, R., \& {Rappaport}, S. 2018, ArXiv e-prints,
  arXiv:1803.03303

\bibitem[{{Winn} {et~al.}(2017){Winn}, {Sanchis-Ojeda}, {Rogers}, {Petigura},
  {Howard}, {Isaacson}, {Marcy}, {Schlaufman}, {Cargile}, \& {Hebb}}]{Winn2017}
{Winn}, J.~N., {Sanchis-Ojeda}, R., {Rogers}, L., {et~al.} 2017, \aj, 154, 60

\bibitem[{{Wurm} {et~al.}(2013){Wurm}, {Trieloff}, \& {Rauer}}]{Wurm}
{Wurm}, G., {Trieloff}, M., \& {Rauer}, H. 2013, \apj, 769, 78

\bibitem[{{Yee} {et~al.}(2017){Yee}, {Petigura}, \& {von Braun}}]{Yee2017}
{Yee}, S.~W., {Petigura}, E.~A., \& {von Braun}, K. 2017, \apj, 836, 77

\bibitem[{{Zeng} {et~al.}(2016){Zeng}, {Sasselov}, \& {Jacobsen}}]{Zeng2016}
{Zeng}, L., {Sasselov}, D.~D., \& {Jacobsen}, S.~B. 2016, \apj, 819, 127

\bibitem[{Zeng {et~al.}(2019)Zeng, Jacobsen, Sasselov, Petaev, Vanderburg,
  Lopez-Morales, Perez-Mercader, Mattsson, Li, Heising, Bonomo, Damasso,
  Berger, Cao, Levi, \& Wordsworth}]{Zeng2019}
Zeng, L., Jacobsen, S.~B., Sasselov, D.~D., {et~al.} 2019, Proceedings of the
  National Academy of Sciences, 116, 9723.
\newblock \url{https://www.pnas.org/content/116/20/9723}

\end{thebibliography}

\begin{rotatetable}
\begin{deluxetable*}{cccccccc}
\tabletypesize{\scriptsize}
\tablecaption{Stellar Parameters \label{tab:stellar}}
\tablehead{
\colhead{System} & \colhead{$T_{\rm eff}$}(K) & \colhead{log~$g$} &\colhead{[Fe/H]} & \colhead{Spectroscopic Source}&\colhead{$M_\star$ ($M_\odot$)}  &\colhead{$R_\star$ ($R_\odot$)}& \colhead{$\rho_\star$ (g cm$^{-3}$)}  }
\startdata
55 Cnc e &5172$\pm$18 &4.43$\pm$0.02 &0.35$\pm$0.10& \citet{Yee2017} & 0.873$^{+0.051}_{-0.035}$& 0.954$^{+0.017}_{-0.018}$& 1.41$^{+0.15}_{-0.10}$\\
CoRot-7b &5313$\pm$73 &4.54$\pm$0.04& 0.03$\pm$0.07& \citet{Torres2012} & 0.884$^{+0.029}_{-0.032}$& 0.8389$^{+0.0093}_{-0.0088}$& 2.11$^{+0.17}_{-0.19}$ \\
GJ 9827b& 4255$\pm$110 & 4.70$\pm$0.15& -0.28$\pm$0.12 & \citet{Niraula2017} &0.606$^{+0.020}_{-0.015}$& 0.5994$^{+0.0081}_{-0.0085}$& 3.97$^{+0.20}_{-0.15}$ \\
HD 3167b& 5261$\pm$60&4.47$\pm$0.05 &0.04$\pm$0.05 & \citet{Christiansen2017} &0.837$^{+0.053}_{-0.043}$& 0.880$^{+0.012}_{-0.013}$& 1.73$^{+0.18}_{-0.14}$\\
K2-106b& 5496$\pm$46& 4.42$\pm$0.05 &0.06$\pm$0.03 & \citet{Sinukoff2017} &0.902$^{+0.057}_{-0.046}$& 0.981$^{+0.019}_{-0.018}$& 1.35$^{+0.14}_{-0.12}$\\
K2-131b &5200$\pm$100 &4.62$\pm$0.10 &-0.02$\pm$0.08 & \citet{Dai2017} & 0.793$^{+0.024}_{-0.040}$& 0.744$^{+0.010}_{-0.009}$& 2.71$^{+0.14}_{-0.19}$\\
K2-141b &4599$\pm$79 &4.62$^{+0.02}_{-0.03}$ &-0.06$^{+0.08}_{-0.10}$ & \citet{Malavolta2018} &0.706$^{+0.021}_{-0.015}$& 0.6862$^{+0.0066}_{-0.0080}$& 3.86$^{+0.16}_{-0.18}$ \\
K2-229b & 5185$\pm$32 &4.56$^{+0.03}_{-0.05}$ & -0.06$\pm$0.02 & \citet{Santerne2018} & 0.804 $^{+0.038}_{-0.048}$& 0.781$^{+0.012}_{-0.012}$& 2.38$^{+0.18}_{-0.20}$\\
K2-291b & 5520$\pm$60 &4.50$\pm$0.05 & 0.08$\pm$0.04 & \citet{Kosiarek} & 0.926 $^{+0.046}_{-0.048}$& 0.904$^{+0.012}_{-0.012}$& 1.77$^{+0.13}_{-0.14}$\\
Kepler-10b& 5708$\pm$28 & 4.344$\pm$0.004& -0.15$\pm$0.04 & \citet{Dumusque2014} & 0.905$^{+0.050}_{-0.044}$& 1.075$^{+0.016}_{-0.015}$& 1.029$^{+0.090}_{-0.084}$ \\
Kepler-78b & 5121$\pm$44 & 4.61$\pm$0.06 & -0.08$\pm$0.04 & \citet{Howard2013} & 0.779$^{+0.032}_{-0.046}$& 0.7475$^{+0.0077}_{-0.0078}$& 2.63$^{+0.16}_{-0.19}$\\
WASP-47e& 5552$\pm$75& 4.34$\pm$0.03 &0.38$\pm$0.05 & \citet{Vanderburg2017} &1.008$^{+0.053}_{-0.047}$& 1.125$^{+0.027}_{-0.026}$& 1.00$^{+0.10}_{-0.09}$ \\
\enddata
\end{deluxetable*}
\end{rotatetable}

\movetabledown=2.3in
\begin{rotatetable}
\begin{deluxetable*}{llllllllll}[p]
\tabletypesize{\scriptsize}
\tablecaption{Planetary Parameters \label{tab:stellar}}
\tablehead{
\colhead{System} & \colhead{$P_{\rm orb}$(days)}& \colhead{$K$(m$~$s$^{-1}$)} & \colhead{$R_p/R_\star$} & \colhead{$M_p$ ($M_\oplus$)} &\colhead{$R_p$ ($R_\oplus$)} & $\rho_p$ (g cm$^{-3}$) &\colhead{Core Mass Fraction$^{1}$} &Water Content$^{2}$ & \colhead{Radial Velocity Source}}
\startdata
55 Cnc e &0.737 & 6.00$^{+0.17}_{-0.17}$ & 0.01821$^{+0.00027}_{-0.00029}$  &7.74$^{+0.37}_{-0.30}$ & 1.897$^{+0.044}_{-0.046}$ &6.25$^{+0.74}_{-0.70}$&  -0.10$\pm$0.14 &$<0.28$  &\citet{Bourrier}\\
CoRoT-7b  & 0.854 &3.37$^{+0.84}_{-0.86}$ & 0.0172$^{+0.0010}_{-0.0011}$ & 4.6$^{+1.1}_{-1.2}$  &1.58$^{+0.10}_{-0.10}$ & 6.5$^{+2.8}_{-3.0}$&  0.11$\pm$0.45 &<0.36 &\citet{Haywood}\\
GJ-9827b &1.209  &4.10$^{+0.35}_{-0.34}$ & 0.02401$^{+0.00038}_{-0.00049}$ & 4.89$^{+0.43}_{-0.42}$ & 1.571$^{+0.033}_{-0.039}$ & 6.9$^{+1.0}_{-1.1}$&  0.22$\pm$0.15 &<0.12&\citet{Teske,Rice}\\
&&&&&&&& &\citet{Prieto}\\
HD-3167b &0.960  &4.08$^{+0.70}_{-0.69}$ & 0.01692$^{+0.00045}_{-0.00050}$ & 5.59$^{+0.98}_{-0.96}$  &1.626$^{+0.048}_{-0.054}$ &7.2$\pm$1.9 & 0.23$\pm$0.27&<0.16 & \citet{Christiansen2017,Gandolfi}\\
K2-106b &0.571  &6.37$^{+0.60}_{-0.62}$  &0.01598$^{+0.00056}_{-0.00057}$ & 7.72$^{+0.80}_{-0.79}$  &1.712$^{+0.068}_{-0.068}$  &8.5$\pm$1.9& 0.40$\pm$0.23&<0.17 & \citet{Sinukoff2017,Guenther}\\
K2-131b  &0.369  &6.55$^{+1.48}_{-1.48}$ & 0.02032$^{+0.00076}_{-0.00064}$ &  6.3$^{+1.4}_{-1.4}$ & 1.651$^{+0.065}_{-0.056}$ & 7.7$^{+2.7}_{-2.6}$& 0.31$\pm$0.34 & <0.16&\citet{Dai2017}\\
K2-141b &0.280 &6.36$^{+0.41}_{-0.40}$ & 0.01993$^{+0.00052}_{-0.00042}$  &5.16$^{+0.35}_{-0.34}$  &1.493$^{+0.041}_{-0.035}$ & 8.5$^{+1.3}_{-1.2}$ & 0.53$\pm$0.15 &<0.02 & \citet{Malavolta2018,Barragan}\\
K2-229b &0.584 & 2.20$^{+0.36}_{-0.37}$  &0.01403$^{+0.00048}_{-0.00052}$  &2.49$^{+0.42}_{-0.43}$  &1.197$^{+0.045}_{-0.048}$ &8.0$^{+2.2}_{-2.3}$ & 0.64$\pm$0.26  &<0.01& \citet{Santerne2018}\\
K2-291b &2.225 & 3.31$^{+0.56}_{-0.56}$  &0.01603$^{+0.00031}_{-0.00037}$  &6.4$^{+1.1}_{-1.1}$  &1.582$^{+0.037}_{-0.042}$ &8.9$^{+2.2}_{-2.3}$ & 0.53 $\pm$ 0.35  & <0.14&\citet{Kosiarek}\\
Kepler-10b &0.837  &2.59$^{+0.36}_{-0.38}$  & 0.012684$\pm$0.000041   &3.57$^{+0.51}_{-0.53}$  &1.489$^{+0.023}_{-0.021}$ & 6.0$\pm$1.1& 0.07$\pm$0.21   & <0.17&\citet{Dumusque2014}\\
Kepler-78b &0.355 & 1.89$^{+0.25}_{-0.25}$  &0.01505$^{+0.00016}_{-0.00017}$ & 1.77$^{+0.24}_{-0.25}$  &1.228$^{+0.018}_{-0.019}$ & 5.26$^{+0.94}_{-0.98}$& 0.08$\pm$0.20 &<0.16 &\citet{Howard2013,Pepe} \\
WASP-47e &0.790  &4.76$^{+0.78}_{-0.79}$ & 0.0144$\pm$0.00020  &6.91$^{+0.81}_{-0.83}$  &1.773$^{+0.049}_{-0.048}$  &6.8$\pm$1.4& 0.09$\pm$0.21   & $<0.21$&\citet{Vanderburg2017}\\
\enddata
\tablenotetext{1}{Iron mass fraction assuming a Fe-MgSiO$_3$ 2-layer model.}
\tablenotetext{2}{2-$\sigma$ Upper limit on the water content on top of an Earth-like Core.}
\end{deluxetable*}
\end{rotatetable}

\end{document}